\documentclass[fleqn,usenatbib]{mnras}

\usepackage{mathptmx}

\usepackage[T1]{fontenc}
\usepackage{ae,aecompl}

\usepackage{graphicx}	
\usepackage{amsmath}	
\usepackage{amssymb}	
\usepackage{rotating}
\usepackage{txfonts}
\usepackage{multicol}        

\newcommand{\flux}{\,erg\,s$^{-1}$\,cm$^{-2}$}	

\title[High-energy emission from HD~93129A]{The high-energy emission from HD~93129A near periastron}

\author[S. del Palacio et al.]{
S. del Palacio,$^{1}$\thanks{E-mail: sdelpalacio@iar.unlp.edu.ar (IAR)}
F. Garc\'ia,$^{2}$
D. Altamirano,$^{3}$
R.~H. Barb\'a,$^{9}$
V. Bosch-Ramon,$^{4}$
\newauthor
M. Corcoran,$^{6,8}$
M. De Becker,$^{5}$
K. Hamaguchi,$^{6,7}$
J. Ma{\'\i}z Apell\'aniz,$^{10}$
P. Munar Adrover,$^{4}$
\newauthor
J.~M. Paredes,$^{4}$
G.~E. Romero,$^{1}$
H. Sana,$^{14}$
M. Tavani,$^{11,12}$
and A. ud-Doula$^{13}$
\\
$^{1}$Instituto Argentino de Radioastronom\'{\i}a (CONICET;CICPBA), C.C. No 5, 1894, Villa Elisa, Argentina\\
$^{2}$Kapteyn Astronomical Institute, University of Groningen, P.O. BOX 800, 9700 AV Groningen, The Netherlands\\
$^{3}$School of Physics and Astronomy, University of Southampton Highfield Campus, Southampton SO17 1PS, UK.\\
$^{4}$Institut de Ci\`encies del Cosmos (ICCUB), Universitat de Barcelona, IEEC-UB, Mart\'{\i} i Franqu\`es 1, E-08028 Barcelona, Spain\\
$^{5}$Space sciences, Technologies and Astrophysics Research (STAR) Institute, University of Li\`ege, Quartier Agora, 19c, All\'ee du 6 Ao\^ut, B5c, B-4000 Sart Tilman, Belgium\\
$^{6}$CRESST II and X-ray Astrophysics Laboratory NASA/GSFC, Greenbelt, MD 20771, USA\\
$^{7}$Department of Physics, University of Maryland, Baltimore County, 1000 Hilltop Circle, Baltimore, MD 21250, USA\\
$^{8}$Institute for Astrophysics and Computational Sciences, Department of Physics, The Catholic University of America, Washington, DC 20064, USA\\
$^{9}$Departamento de Astronom\'ia, Universidad de La Serena, Av. Juan Cisternas 1200 Norte, La Serena, Chile\\
$^{10}$Centro de Astrobiolog\'ia, CSIC-INTA, Campus ESAC. C. Bajo del Castillo s/n, 28 692, Villanueva de la Ca\~nada, Madrid, Spain\\
$^{11}$INAF-IAPS, via del Fosso del Cavaliere 100, I-00133 Roma, Italy\\
$^{12}$Dip. di Fisica, Univ. di Roma "Tor Vergata", via della Ricerca Scientifica 1, I-00133 Roma, Italy\\
$^{13}$Dept. of Physics, Penn State Scranton, 120 Ridge View Drive, Dunmore, PA 18512, USA\\
$^{14}$Instituut voor Sterrenkunde, KU Leuven, Celestijnenlaan 200D, 3001, Leuven, Belgium
}

\date{Accepted XXX. Received YYY; in original form ZZZ}

\pubyear{2019}

\begin{document}
\label{firstpage}
\pagerange{\pageref{firstpage}--\pageref{lastpage}}
\maketitle

\begin{abstract}
   We conducted an observational campaign towards one of the most massive and luminous colliding wind binaries in the Galaxy, HD~93129A, close to its periastron passage in 2018. During this time the source was predicted to be in its maximum of high-energy emission. Here we present our data analysis from the X-ray satellites \textit{Chandra} and \textit{NuSTAR} and the $\gamma$-ray satellite \textit{AGILE}.
   High-energy emission coincident with HD~93129A was detected in the X-ray band up to $\sim$18~keV, whereas in the $\gamma$-ray band only upper limits were obtained. We interpret the derived fluxes using a non-thermal radiative model for the wind-collision region. We establish a conservative upper limit for the fraction of the wind kinetic power that is converted into relativistic electron acceleration, $f_\mathrm{NT,e} < 0.02$. In addition, we set a lower limit for the magnetic field in the wind-collision region as $B_\mathrm{WCR} > 0.3$~G. We also argue a putative interpretation of the emission from which we estimate $f_\mathrm{NT,e} \approx 0.006$ and $B_\mathrm{WCR} \approx 0.5$~G.
   We conclude that multi-wavelength, dedicated observing campaigns during carefully selected epochs are a powerful tool for characterising the relativistic particle content and magnetic field intensity in colliding wind binaries.
\end{abstract}




%
\section{Introduction}\label{intro}
%

The role of massive stars as accelerators of Galactic cosmic rays has been gaining more interest in the recent years \citep[e.g.][]{Seo2018,Aharonian2018, Prajapati2019}. Unfortunately, it is not possible to directly assess the efficiency of cosmic ray acceleration from observations. The study of massive colliding-wind binaries (CWBs) is key in understanding the non-thermal physics taking place in systems harbouring massive stars \citep{DeBecker2017}. The signature of relativistic particles is a non-thermal spectrum, typically a power-law with additional features such as a spectral break or an exponential cutoff. In the radio band, non-thermal emission can be produced by relativistic electrons interacting with the magnetic field \citep{Eichler1993}. The detection of this synchrotron radiation in radio observations of CWBs have provided conclusive evidence that at least dozens of CWBs in the Galaxy are capable of accelerating cosmic rays \citep{DeBecker2013}. The most accepted scenario is that relativistic particles accelerate via diffusive shock acceleration in the strong shocks of the wind-collision region \citep[WCR;][]{Pittard2006B}. However, the cosmic ray acceleration efficiency cannot be proven from radio data alone \citep[e.g.][an references therein]{delPalacio2016}. Therefore, additional information of the high-energy spectrum of CWBs is required in order to make progress in the characterisation of their non-thermal physics.

The idea that CWBs can produce significant non-thermal emission at high energies (X-rays and $\gamma$-rays) is supported by the detection of such a radiation from the exceptional binary $\eta$-Carinae \citep{Tavani2009, Reitberger2015, Hamaguchi2018, HESS2020}, although this fascinating object is hardly representative for CWBs in general. Another CWB, $\gamma^2$~Vel, has also been likely detected at $\gamma$-rays \citep{Pshirkov2016, Marti-Devesa2020}. However, none of these two systems has been detected as a non-thermal radio emitter, most likely due to synchrotron emission being self-absorbed in the system \citep[e.g.][]{Benaglia2019}. If more CWBs are confirmed as non-thermal emitters at high energies, it will be possible to assess the role of CWBs as cosmic ray injectors. Moreover, it will give compelling support to the idea that massive binary systems are frequently $\gamma$-ray emitters \citep{Benaglia2003, DeBecker2017}. 

In this work, we focus on the study of the high-energy emission from the extreme binary HD~93129A. This system, located in the core of the young star cluster Trumpler~14, is the most massive un-evolved binary in the solar vicinity. It is made up by two O2 stars that are among the earliest, hottest, most luminous, and with highest mass-loss rates in the Galaxy \citep{Walborn2002, Maiz2008}. The most recent published ephemeris \citep{Maiz2017} yielded a binary period of $\sim 120$~years, a periastron passage in $2018.54^{+0.54}_{-0.32}$, and an orbital inclination $i \sim 99\degr$ (i.e., nearly edge-on). We have kept obtaining VLTI astrometry of the HD~93\,129~Aa,Ab visual pair after the orbital solution that was published in \citet{Maiz2017}. 
The new data includes three post-periastron observations that help to significantly reduce the uncertainties in the orbital determination. Regarding the most relevant parameters for this study, the time of periastron passage is now tightly constrained to $2018.70^{+0.22}_{-0.12}$ (i.e., a 1-$\sigma$ uncertainty of just 2 months), and the distance at periastron is also well constrained within a $\sim 5\%$ uncertainty, namely as $7.91\pm 0.42$~mas ($18.6\pm 1.0$~AU). In addition, the non-thermal synchrotron radiation from the WCR has been resolved in radio by \citet{Benaglia2015}, allowing for a partial characterisation of the relativistic particle energy distribution, and confirming that particle acceleration occurs in this binary. 

We conducted our high-energy observational campaign towards this system near its periastron passage. The high-energy non-thermal emission from the WCR, produced by inverse Compton scattering of stellar photons, was expected to peak towards periastron due to the increased number of target stellar photons.
The non-thermal emission from this system was predicted by \citet{delPalacio2016} using a broadband radiative model for the WCR that takes into account (i) the evolution of the accelerated particles streaming along the shocked region, (ii) the emission by different radiative processes, and (iii) the attenuation of the emission propagating through the local matter and radiation fields. On the basis of their analysis, \citet{delPalacio2016} suggested that HD~93129A was a promising candidate for detecting high-energy emission close to its periastron passage. A multi-wavelength observational campaign was carried out to get a complete picture of this event. In this work, we present the analysis of high-energy observations performed with \textit{Chandra} (soft X-rays, 0.3--10~keV), \textit{NuSTAR} (hard X-rays, 3--79~keV) and \textit{AGILE} ($\gamma$-rays, 50~MeV--300~GeV) close to periastron.

The paper is organised as follows. The description of the observations is given in Sect.~\ref{sec:obs}, 
and the procedure adopted to reduce and analyse each data set in Sect.~\ref{sec:spectral_fitting}. We present the results from this analysis in Sect~\ref{sec:results}. In Sect.~\ref{sec:disc} we discuss the constraints of our results in the context of theoretical models, and we present a summary and concluding remarks in Sect~\ref{sec:conc}.

%
\section{Observations}\label{sec:obs}
%

We observed the binary HD~93129A close to its periastron passage in 2018. A summary of the analysed observations is presented in Table~\ref{tab:observations}. In the following subsections we describe them in more detail.
\begin{table*}
\caption{Summary of the X-ray observations analysed.}
\label{tab:observations}
\begin{tabular}{l c c c c c c c}
\hline
Instrument      &   Obs.~ID     & Date (MJD)&   Exposure time (ks) &  Effective time (ks) \\
\hline
\textit{Chandra}&   4495        & 53269.7   &   58.06     & 57.33 \\ 
\textit{Chandra}&   20152       & 58288.4   &   25.06     & 23.84 \\
\textit{Chandra}&   20153       & 58343.1   &   25.06     & 23.84 \\
\hline
\textit{NuSTAR} &   30402001002 & 58290.76  &   27.59     & 27.40 \\
\textit{NuSTAR} &   30402001004 & 58350.03  &   33.07     & 32.20 \\
\hline
\end{tabular}
\end{table*}

\subsection{\textit{Chandra} observations}\label{sec:chandra}

\begin{figure}
\includegraphics[width=1.0\linewidth]{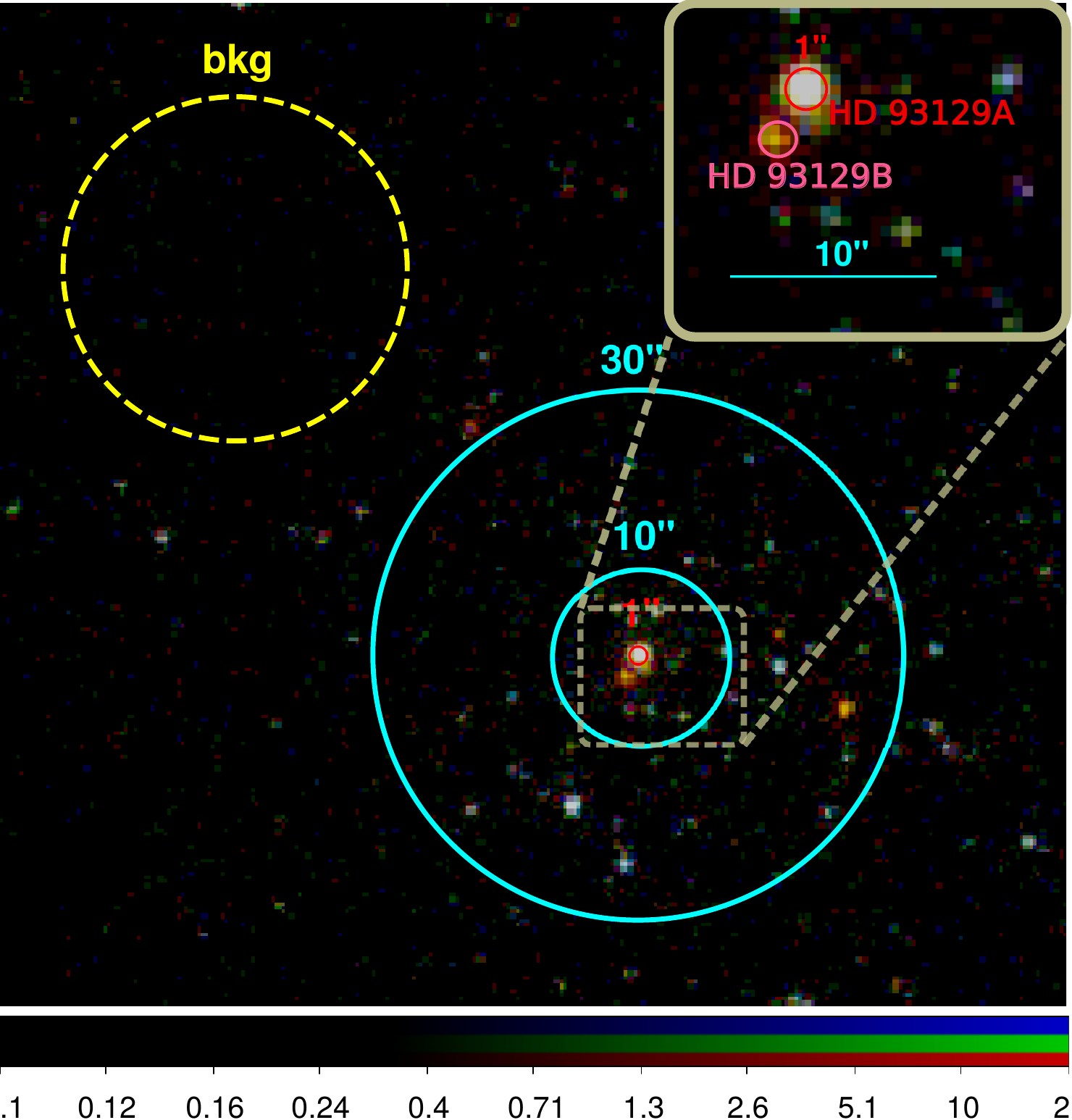}
\caption{\textit{Chandra} RGB image from obs. 20152 (red is 0.5--1.2~keV, green is 1.2--2.5~keV, and blue is 2.5--7~keV). Circular regions centred at the position of HD~93129A and of different radii are shown for reference, as well as the selected background extraction region. The position of the source HD~93129B (not part of the system HD~93129A) is also marked for clarity.}
\label{fig:chandra_img}
\end{figure}

The \textit{Chandra X-Ray Observatory} is unique in terms of the high angular resolution that it reaches in the X-ray band. The ACIS instrument on board of the satellite is capable of focusing X-rays with energies in the range 0.3--10~keV with subarcsecond resolution  \citep{Chandra2002}. This is particularly useful for our study as it allows us to quantify the contamination from background sources such as the Trumpler~14 association. An image of the field of vew of HD~93129A is shown in Fig.~\ref{fig:chandra_img}.

Since we are interested in studying the continuum emission from HD~93129A (i.e., not the detailed spectral lines), we restrict our analysis to on-axis observations without gratings. We found one archival observation (Obs.~id. 4495) from 2004 with 58~ks exposure taken with ACIS-I. In addition, we have observations from June and August 2018 (Obs.~id. 20152 and 20153, respectively), both with 25~ks exposure taken with ACIS-S.

We reduced the \textit{Chandra} data using \texttt{CIAO} v.4.11 \citep{Fruscione2006}. For all the data sets we used the \texttt{CIAO} task \texttt{chandra\_repro} to reprocess the data using the latest calibration files available (CALDB~v.4.8.4.1).

The source spectra were extracted using the task \texttt{specextract} with the options appropriate for point source analysis (\texttt{weight=no} and \texttt{correctpsf=yes}). When considering a larger region ($10\arcsec$ or $30\arcsec$) to estimate contamination by nearby sources, \texttt{weight} was set to \texttt{yes}. We found little variation between both observations from 2018 (less than 2\% in the integrated flux above 3~keV), so the task \texttt{combine\_spectra} was used to combine their spectra. Finally, we used the task \texttt{ftgrouppha} from \texttt{HEASOFT~v.6.26} to group the spectra with the optimal binning scheme by \citet{Kaastra2016}.

\subsection{\textit{NuSTAR} observations}\label{sec:nustar}

The \textit{NuSTAR} X-ray observatory, launched in 2012, counts with two co-aligned hard X-ray grazing incidence telescopes labelled by their focal plane modules FPMA and FPMB. These instruments are capable of observing in the 3--79 keV energy range with an angular resolution of 18\arcsec 
 \citep[half power diameter of 58\arcsec;][]{NuSTAR2013}.
 
\textit{NuSTAR} observations with 30~ks exposure were performed within a week of the \textit{Chandra} observations. This makes both data sets almost simultaneous considering the expected variability timescale.
The \texttt{nupipeline} task was used to create level 2 data products. We used the option \texttt{saacalc=2 saamode=optimized tentacle=yes} to filter high background epochs. This leads to $< 3\%$ data loss\footnote{\url{http://www.srl.caltech.edu/NuSTAR_Public/NuSTAROperationSite/SAA_Filtering/SAA_Filter.php}.}. We used the \texttt{nuproducts} task to create level 3 data products. We present an image (3--11~keV; $PI$ channels 35--235) in Fig.~\ref{fig:nustar_img}, showing the selected source and background extraction regions for each observation.

\subsection{\textit{AGILE} observations}\label{sec:agile}

We analysed the data collected between 01/01/2018 and 31/12/2018 
by the Gamma-Ray Imaging Detector \citep[GRID;][]{Barbiellini2002, Prest2003} on board the \textit{AGILE} satellite  \citep[see][for a detailed description of the \textit{AGILE} payload]{Tavani2009a}. 
The \textit{AGILE}-GRID is sensitive in the 30 MeV--50 GeV energy band. The point spread function (PSF) at 100 MeV and 400 MeV is $4.2\degr$ and $1.2\degr$ (68\% containment radius), respectively \citep{Sabatini2015}. We restricted our analysis to photon energies from 100~MeV to 50~GeV. The angular resolution of \textit{AGILE} in this range is $0.9\degr$, so that significant contamination from sources close to HD~93129A is expected.

Since 2009 \textit{AGILE} observes in `spinning' mode, covering a large fraction of the sky with a controlled rotation of the pointing axis. In this observing mode, typical two-day integration-time sensitivity for sources in the Galactic plane and photon energy above 100~MeV is $\sim 10^{-6}$ photons~cm$^{-2}$~s$^{-1}$ ($3\sigma$).

The analysis of the \textit{AGILE}-GRID data was carried out with the new \texttt{Build 23} scientific software, \texttt{FM3.119} calibrated filter and \texttt{I0025} response matrices. The consolidated archive, available from the ASI Data Center (\texttt{ASDCSTDk}), was analyzed by applying South Atlantic Anomaly event cuts and $80\degr$ Earth albedo filtering. Only incoming $\gamma$-ray events with an off-axis angle lower than $60\degr$ were selected for the analysis. Statistical significance and flux determination of the point sources were calculated by using the \textit{AGILE} multi-source likelihood analysis (MSLA) software \citep{Bulgarelli2012} based on the Test Statistic (TS) method as formulated by \citet{Mattox1996}. This statistical approach provides a detection significance assessment of a $\gamma$-ray source by comparing maximum-likelihood values of the null hypothesis (no source in the model) with the alternative hypothesis (point source in the field model). In this work we report 68\% confidence level flux upper limits if TS $< 9$ (detection significance $<3$). 
To estimate the likelihood of a detection, two different considerations were made to account for multiple nearby sources: i) excluding $\eta$-Car, and ii) including $\eta$-Car. For each of these background models we constructed weekly and monthly binned light-curves.

\section{Spectral fitting}\label{sec:spectral_fitting}

The X-ray spectrum of a CWB has many physical components produced by the two individual stellar winds and the two shocked winds. The individual stellar winds produce only low energy X-rays \citep[typically $<1$~keV,][]{Owocki1988}, whereas the shocked winds can produce hard X-rays above 3~keV via thermal emission or by IC scattering of stellar photons by relativistic electrons \citep{Hamaguchi2018}. These emission components are affected by absorption by the intertellar medium and within the stellar winds \citep{Cohen2011}. This means that a realistic and physically consistent model requires several components, which unfortunately leads to a degeneracy when fitting real data. In practice, the observed spectrum can be well approximated with a two-temperature \textit{apec}\footnote{http://www.atomdb.org/} component together with a \textit{phabs} \citep{Anders1989} absorption component \citep{Pittard2010II}. The \textit{apec} is an optically thin plasma thermal emission model that has three parameters: the plasma temperature ($kT$), normalisation ($N$), and abundance ($A$). This model is adequate to represent the X-ray emission from a hot plasma such as those produced by hot stellar winds of massive stars \citep{FeldX,Owocki2013} and CWBs \citep{Stevens1992,PP2010,Hamaguchi2018}. The low temperature \textit{apec} includes the thermal emission from the individual stellar winds and the outer (colder) regions of the WCR. The high-temperature \textit{apec} represents the thermal emission from near the apex of the WCR. In addition, a multiplicative \textit{phabs} model is used as an effective absorption component that takes into account the attenuation in the ISM and the stellar winds. We note that this model, despite being an oversimplification, is sufficient to qualitatively characterise the spectrum of the source (see Sect.~\ref{sec:res-chandra}). An additional power-law component can also be introduced to account for putative non-thermal emission. Finally, we stress that this study is focused on the high-energy emission, so a detailed spectral fitting of low energy photons ($< 3$~keV) will be addressed in a separate work, together with a detailed analysis of the temporal variability of such thermal emission.

The spectral analysis was done using the software \texttt{XSPEC~v.12.10.1f} \citep{xspec1996,xspec2001}. The C-stat minimisation approach is used to check the adequacy of the fit, given that the source is not very bright in hard X-rays. Error-bars are calculated at $1\sigma$ for all cases.

%
\section{Results}\label{sec:results}
%


%
\subsection{\textit{Chandra}}\label{sec:res-chandra}

We aim to characterise the X-ray spectrum of HD~93129A and its environment using the \textit{Chandra} data. In Fig.~\ref{fig:chandra_img} we show the \textit{Chandra} field of view centred at the position of HD~93129A. The source is resolved thanks to \textit{Chandra}'s high angular resolution ($\approx 0.5\arcsec$ on axis). Several sources are detected within $10\arcsec$ from HD~93129A. We therefore extract the spectrum from HD~93129A using a $1\arcsec$ circular region so that the background/foreground contamination is minimal. The PSF is such that it ensures that almost 100\% of the photons with $E<1$~keV, and $\sim 85\%$ for those with $E \sim 10$~keV, are captured \citep{Weisskopf2002}.

\begin{figure}
\includegraphics[width=1.0\linewidth]{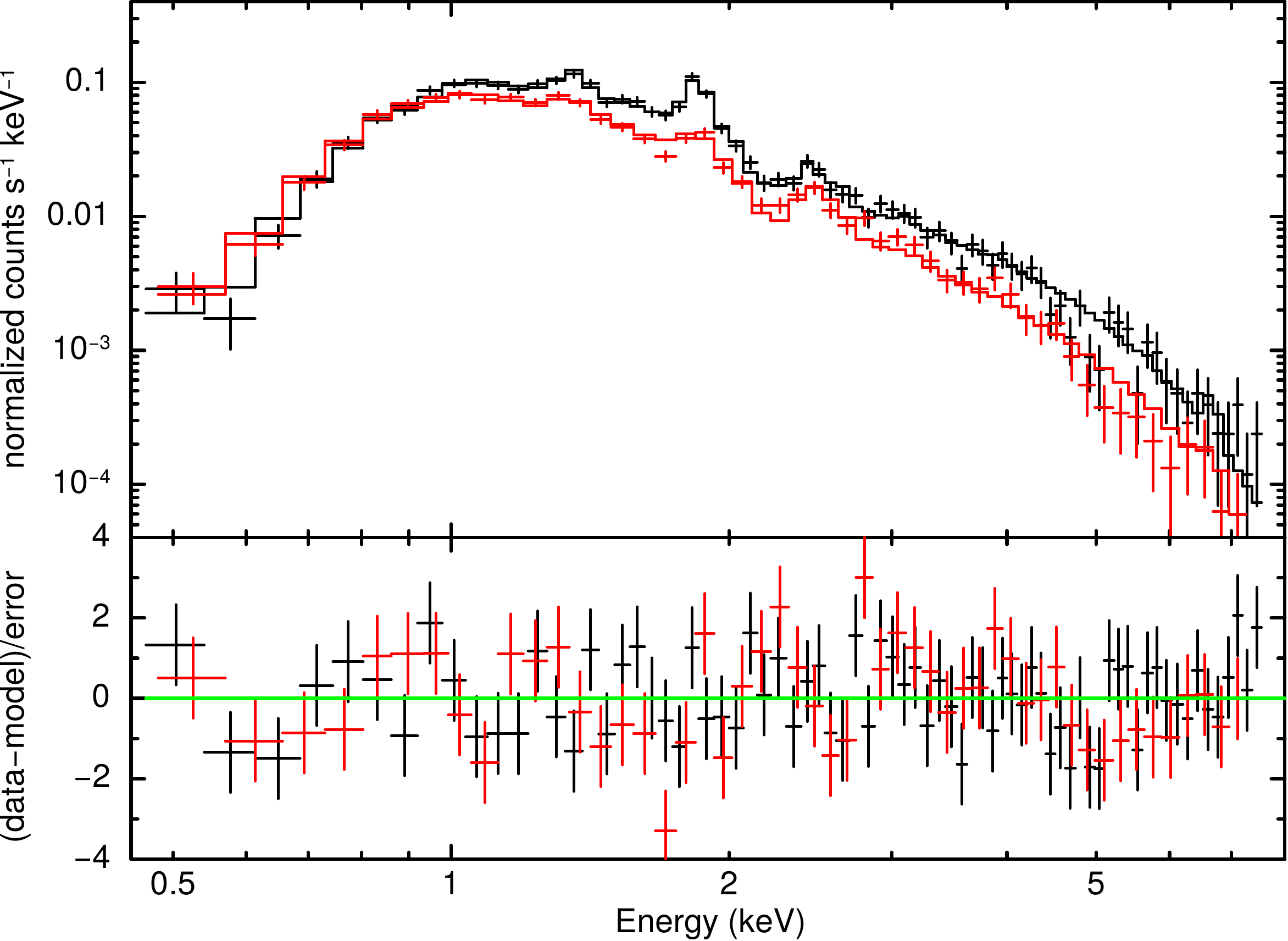}
\caption{Spectra of HD~93129A as observed with \textit{Chandra} in 2018 (black) and 2004 (red) extracted from a $1\arcsec$ circular region. The best-fit model ([\textit{phabs*(vapec$_1$+vapec$_2$)}]) is also shown, with the parameters detailed in Table~\ref{tab:fit_parameters}. }
\label{fig:chandra_spec}
\end{figure}

We compare the spectra from the 2004 and 2018 observations in Fig.~\ref{fig:chandra_spec}.
The source was detected up to 8~keV. We fitted both spectra simultaneously with a model [\textit{phabs*(vapec$_1$+vapec$_2$)}]. We considered the same element abundances for both \textit{vapec} components. We defined a common abundance $A$ for all elements except Si and S, which had prominent emission lines in the spectra, and Fe, which had a significantly lower abundance than the rest. We tried to minimise the amount of free parameters to fit while still obtaining a good spectral fit. For that reason, we tied the abundances of $A$ and Fe between epochs, and allowed only S and Si abundances to vary. The temperature of the colder \textit{vapec} component, $T_1$, was also tied between epochs. The values of $N_{\rm H}$, $T_2$, and normalisations $norm_1$ and $norm_2$, were fitted independently for each epoch. The best fit values are shown in Table~\ref{tab:fit_parameters}. We found that between 2004 and 2018 the observed flux increased by more than 50\% in the 0.5--3~keV range and 100\% in the 3--8~keV range. 

\setlength{\tabcolsep}{3.1pt}
\begin{table}
\caption{Results of the fitting of \textit{Chandra} spectra of HD~93129A using a [\textit{phabs*(vapec+vapec)}] thermal model. C-statistics were used. The errors at 1-$\sigma$ level are specified for all parameters and for the observed 
flux.}
\label{tab:fit_parameters}
\centering
\begin{tabular}{l c c c c}
\hline \hline
 & \multicolumn{2}{c}{\underline{$1\arcsec$ extraction region}} & \multicolumn{2}{c}{\underline{$30\arcsec$ extraction region}} \\[2pt]
 Parameter   & 2004            & 2018       &   2004    &     2018     \\
\hline
$N_{\rm H}$ (10$^{22}$\,cm$^{-2}$)& $0.69^{+0.06}_{-0.05}$ &  $0.73^{+0.06}_{-0.05}$ &  $0.64^{+0.04}_{-0.03}$ & $0.60^{+0.04}_{-0.03}$ \\[2pt] 
$A$                         &  \multicolumn{2}{c}{$0.17^{+0.04}_{-0.03}$}  &  \multicolumn{2}{c}{$0.13^{+0.02}_{-0.02}$} \\[2pt] 
Si                          & $0.48^{+0.09}_{-0.08}$ &  $0.52^{+0.09}_{-0.07}$ &  $0.26^{+0.05}_{-0.04}$ &  $0.39^{+0.11}_{-0.10}$ \\[2pt] 
S                           & $1.13^{+0.26}_{-0.21}$ &  $0.69^{+0.17}_{-0.15}$ &  $0.58^{+0.14}_{-0.13}$ &  $0.36^{+0.11}_{-0.10}$ \\[2pt] 
Fe                          &  \multicolumn{2}{c}{$0.07^{+0.02}_{-0.02}$}  &  \multicolumn{2}{c}{$0.06^{+0.01}_{-0.01}$} \\[2pt] 
$k\,T_1$ (keV)              &  \multicolumn{2}{c}{$0.35^{+0.04}_{-0.03}$}  &  \multicolumn{2}{c}{$0.35^{+0.01}_{-0.02}$} \\[2pt] 
$norm_1$ ($10^{-2}$ cm$^{-5}$) & $1.30^{+0.52}_{-0.30}$ &  $2.56^{+1.02}_{-0.58}$ &  $2.86^{+0.64}_{-0.49}$ &  $3.32^{+0.75}_{-0.55}$ \\[2pt] 
$k\,T_2$ (keV)              & $1.80^{+0.13}_{-0.10}$ &  $2.35^{+0.16}_{-0.14}$ &  $2.41^{+0.14}_{-0.12}$ &  $2.68^{+0.05}_{-0.04}$ \\[2pt] 
$norm_2$ ($10^{-3}$ cm$^{-5}$) & $1.66^{+0.19}_{-0.18}$ &  $2.22^{+0.20}_{-0.19}$ &  $3.48^{+0.23}_{-0.24}$ &  $3.98^{+0.22}_{-0.19}$\\[2pt] 
$F_{0.5-3}$ ($10^{-13}$~erg\,cm$^{-2}$\,s$^{-1}$) & $8.39^{+0.01}_{-0.46}$ &  $13.0^{+0.1}_{-0.8}$ &  $16.9^{+0.1}_{-0.5}$ &  $21.3^{+0.0}_{-0.6}$\\[2pt] 
$F_{3-8}$ ($10^{-13}$~erg\,cm$^{-2}$\,s$^{-1}$)  & $1.85^{+0.12}_{-0.14}$ &  $3.79^{+0.17}_{-0.26}$ &  $5.88^{+0.18}_{-0.26}$ &  $7.82^{+0.13}_{-0.31}$ \\[2pt] 
\hline
$C_\mathrm{stat}$/d.o.f. ($\chi^2_\mathrm{red}$)  & \multicolumn{2}{c}{140.7/110 (1.40)} &  \multicolumn{2}{c}{174.3/124 (1.40)}   \\
\hline
\end{tabular}
\end{table} 

In order to compare the \textit{Chandra} and \textit{NuSTAR} spectra, we need to use a $30\arcsec$ extraction region for both. For this reason, we compared how selecting a $30\arcsec$ extraction region affected the \textit{Chandra} spectrum from 2018. We fitted the \textit{Chandra} spectrum in the 3.0--7.6~keV energy range (the maximum energy is set by the larger background contamination dominating the spectrum, see Fig.~\ref{fig:chandra_1vs30}) for two different extraction regions of radii $1\arcsec$ and $30\arcsec$. We tested whether a complex model was required to fit the spectrum in this energy range. For this we checked that in this energy range the contribution from the low temperature \textit{vapec} and from the absorption \textit{phabs} were negligible. We therefore fitted the spectra with a single high temperature \textit{apec} component. This component has $kT \approx 2.35 \pm 0.25$~keV for both extraction regions. We conclude that there is no significant spectral shape difference in the 3--7.6~keV energy range induced by the increased background contamination in the 2018 \textit{Chandra} spectra extracted from a $30\arcsec$ region with respect to the one from a $1\arcsec$ region.

In addition, we also checked the 2004--2018 variability considering a $30\arcsec$ extraction region. The spectral fitting parameters are summarised in Table~\ref{tab:fit_parameters}. In this case, the flux between the two epochs increased only in $26\%$ in the 0.5--3~keV band and $33\%$ in the 3--8~keV band; similar increase factors were obtained for a $10\arcsec$ region as well (Fig.~\ref{fig:chandra_1vs30}). Interestingly, we can also estimate the average flux of the surrounding sources within an annulus given by the difference between the $30\arcsec$- and the $1\arcsec$-extraction regions. We obtain that the flux variations for the nearby (background) sources between 2004 and 2018 are $< 10\%$ in both of these energy bands (this can be appreciated more clearly in Fig.~\ref{fig:chandra_annulus}). Therefore, the observed flux variability within the $30\arcsec$ region seems to be governed by HD~93129A. We took into account this information when interpreting the \textit{NuSTAR} data in which the binary is not resolved from the surrounding sources (Fig.~\ref{fig:chandra_img}).

%
\subsection{\textit{NuSTAR}} \label{sec:res-nustar}

\begin{figure}
\begin{center}
\includegraphics[angle=270,width=1.0\linewidth]{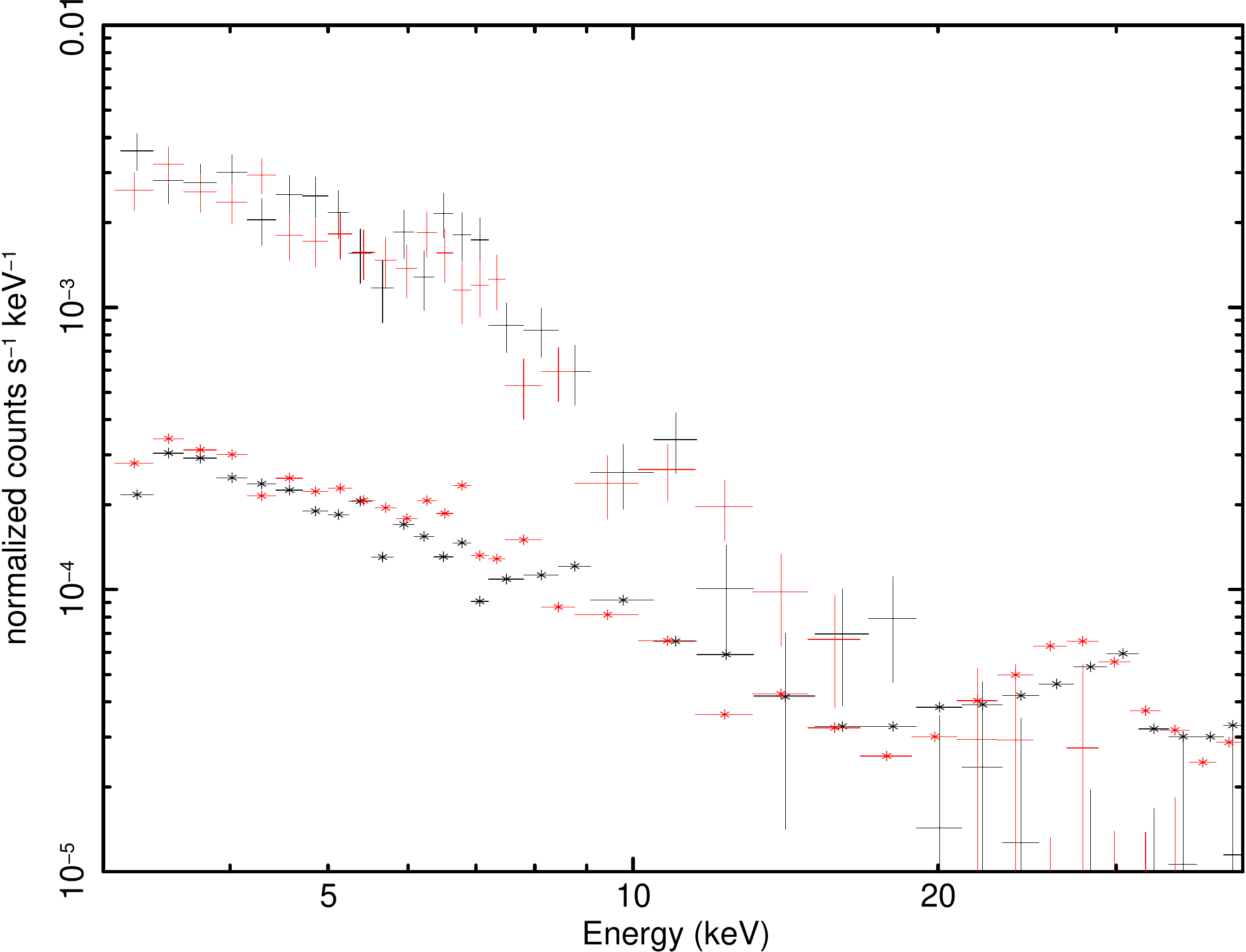}\\
\includegraphics[width=1.0\linewidth]{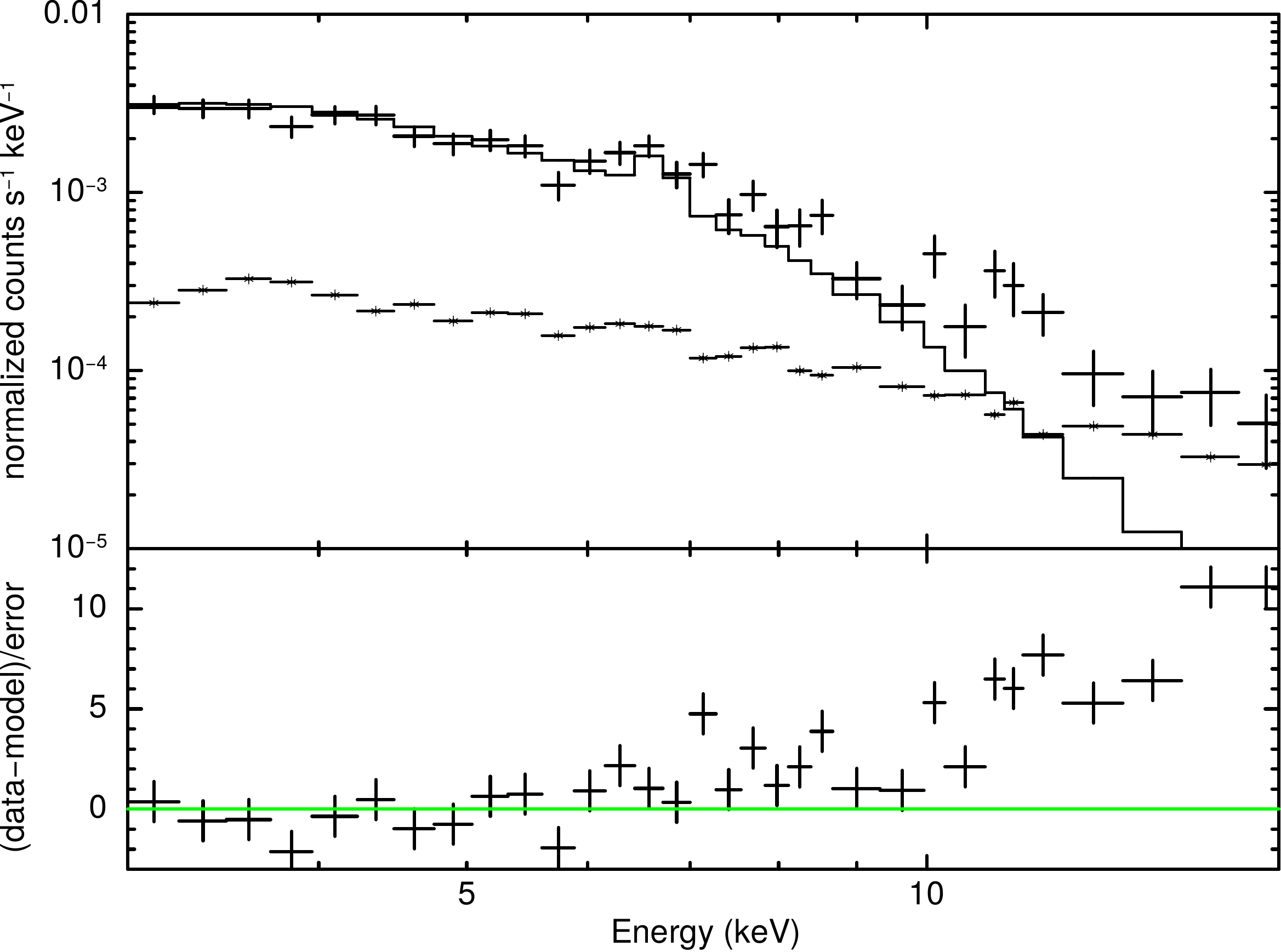}
\end{center}
\caption{Top: \textit{NuSTAR} source and background spectra for each observation (first observation is black, second one is red) in the 3--40~keV energy range. 
The cameras at each observation epoch were co-added with \texttt{addspec}, grouped with \texttt{ftgroup} and rebinned in \texttt{XSPEC} with \texttt{rebin=4}.
Bottom: \textit{NuSTAR} source and background spectra obtained by co-adding both observations, shown in the 3--20~keV energy range. The model overlayed is the one that fits the \textit{Chandra} spectrum in the 3--7.5~keV range, with the residuals shown in the bottom panel. The model under-predicts the emission above 8~keV, which can be explained by the presence of either a hotter component or a non-thermal component.} 
\label{fig:nustar_spec}
\end{figure}

The 3--18~keV fluxes between both FPMA and FPMB cameras in each observation were consistent within less than 5\%. We therefore co-added both cameras using \texttt{addspec}. In Fig.~\ref{fig:nustar_spec} we compare the source and background spectra grouped with \texttt{ftgroup}. The source is detected above background with a high significance at $\approx 13$~keV ($4\sigma$ in the 11.3--15.1~keV energy range), and with a lower significance at $\approx 18$~keV ($1.7\sigma$ in the 15.2--17.8~keV energy range). 
We confirmed this result by selecting different background extraction regions, 
although the total flux in 3--10~keV can vary within $10\%$ depending on the selected background. Moreover, the fluxes between the two observations differ only by $12\%$, which is almost within 1-$\sigma$ level and can be attributed to calibration uncertainties \citep[up to 10\%;][]{Madsen2015} and a different background level (see top panel from Fig.~\ref{fig:nustar_spec}). We therefore assume that the whole $\textit{NuSTAR}$ data set is comparable and co-add both observing epochs with \texttt{addspec}.

In Fig.~\ref{fig:nustar_spec} we plotted together the X-ray data from \textit{Chandra} and \textit{NuSTAR}. The spectra seemed to match well up to 6--7~keV. To check this, we fitted the \textit{NuSTAR} spectrum below 7~keV with a model, and the \textit{Chandra} spectrum from the 30\arcsec-region with the same model times a constant. The fitted value of the constant is $1.036^{+0.056}_{-0.050}$, which is fully consistent with not needing a re-scaling for the data sets (i.e., setting the constant to one). We found that the model fitted using only the \textit{Chandra} data matches very well the \textit{NuSTAR} spectrum up to 6--7~keV. However, the \textit{Chandra} model underestimates the emission above 7~keV. This could be simply due to a poorly constrained high-energy spectrum in the \textit{Chandra} data resulting in a bad extrapolation of the fitted model. Regardless of its origin, the fact that this model falls below the observed flux in the \textit{NuSTAR} spectrum suggests the presence of an additional component, either thermal and hotter or non-thermal. 

\begin{figure}
\includegraphics[width=1.0\linewidth]{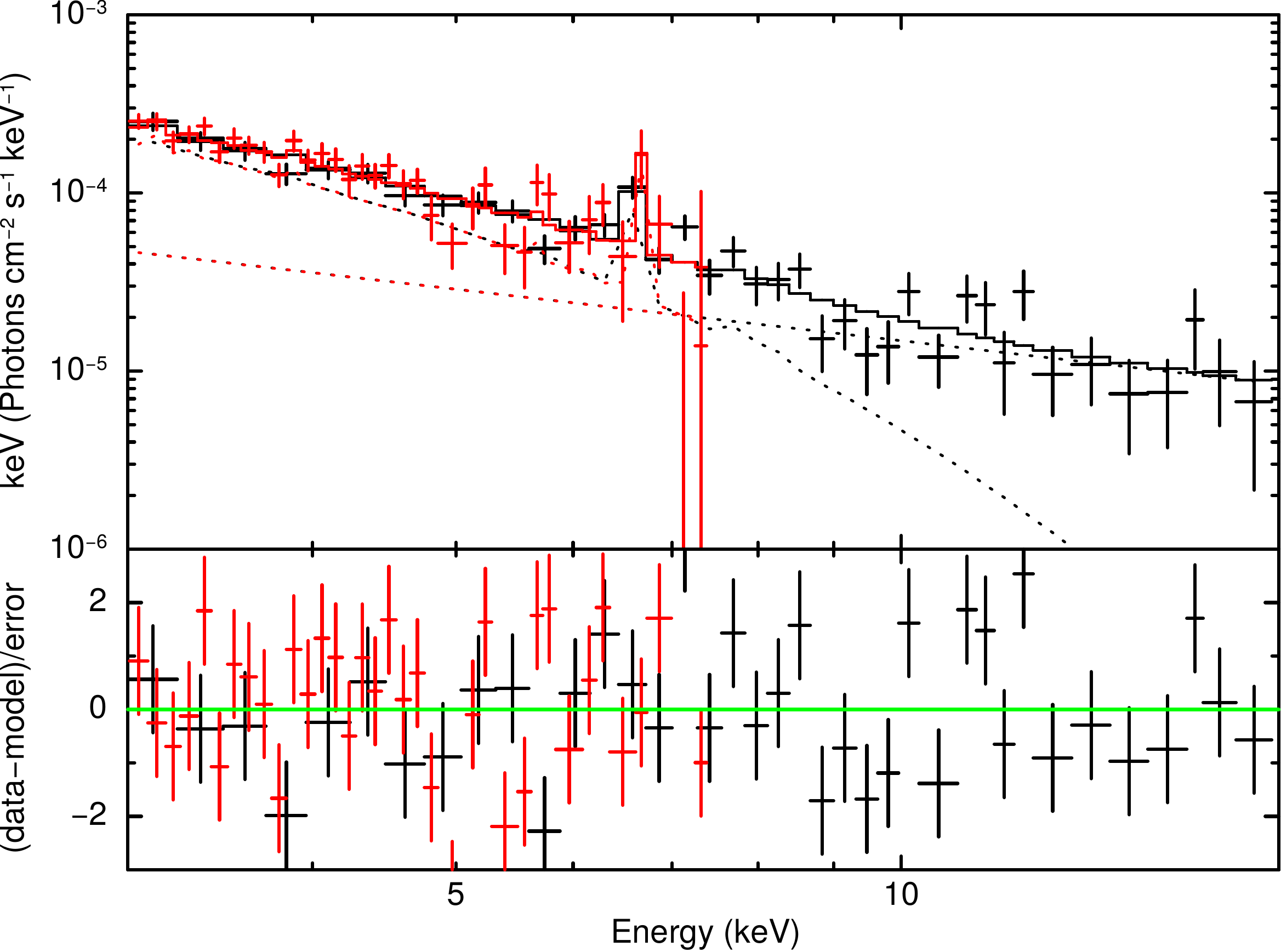}
\caption{\textit{Chandra} (red) and \textit{NuSTAR} (black) spectra for 2018. The fitted model is a [\textit{apec + power-law}], with $kT \approx 2.1$~keV and $\Gamma \approx 1.9$ (see text for details). }
\label{fig:nustar_chandra_spec}
\end{figure}

In Fig.~\ref{fig:nustar_chandra_spec} we show a fit to the \textit{Chandra} and \textit{NuSTAR} spectra using an [\textit{apec + power-law}] model. This fit yields $kT = 2.1 \pm 0.2$~keV, $A = 0.21$ and $\Gamma \approx 2.0^{+0.5}_{-0.7}$ ($C=92.6$, with 87 d.o.f., and ${\chi^2}_\mathrm{red} = 1.11$). We use \textit{cflux} to obtain the unabsorbed flux from the model and obtain $F_{3-8} = 7.84^{+0.35}_{-2.61} \times 10^{-13}$\flux\, and $F_{8-18} = 2.35^{+0.20}_{-0.21} \times 10^{-13}$\flux. We also calculate the individual unabsorbed flux of the power law component and obtain $F_{8-18} = 1.93^{+0.26}_{-0.36} \times 10^{-13}$\flux. 

We also considered the possibility of having only thermal emission. We fitted the \textit{Chandra} + \textit{NuSTAR} data with an [\textit{apec + apec}] model, which yields $A = 0.24$ (tied for both components), $kT_1 = 1.7^{+0.4}_{-0.2}$~keV and $kT_2 = 8.6^{+13}_{-1.5}$~keV ($C=89.8$, with 80 d.o.f., and ${\chi^2}_\mathrm{red} = 1.13$). The unabsorbed fluxes in this case are $F_{3-8} = 7.7^{+0.26}_{-0.24} \times 10^{-13}$\flux and $F_{8-18} = 2.30^{+0.18}_{-0.18} \times 10^{-13}$\flux. 

%
\subsection{\textit{AGILE}}\label{gamma}

The result from the analysis of the \textit{AGILE} data is consistent with an upper limit to the flux in 50~MeV--100~GeV of $2\times10^{-7}$~cm$^{-2}$~s$^{-1}$. 
This is confirmed by the light-curve obtained for HD~93129A using both a weekly and a monthly binning and including the background emission from multiple sources nearby. An example light curve is shown in Fig.~\ref{fig:agile}.
We note that there is a low significance detection of emission near April 2018 at a 4-$\sigma$ level if emission from $\eta$-Car is not considered. However, it is not likely related with HD~93129A, as discussed in Sect.~\ref{sec:disc}.

%
\section{Discussion}\label{sec:disc}
%

In Sect.~\ref{sec:results} we presented the results from the observations with \textit{Chandra}, \textit{NuSTAR} and \textit{AGILE}. The source HD~93129A was clearly detected in the X-ray domain ($0.5-18$~keV range), whereas in the $\gamma$-ray domain no significant emission was detected. 

\subsection{Observations versus predictions}\label{sec:obs_vs_predictions}

From the analysis of the \textit{Chandra} data we can conclude that there is flux variability between 2004 and 2018.
This variability is more clearly seen when a small $1\arcsec$ extraction region is considered. For larger extraction regions with size $\geq 10\arcsec$ the background level becomes comparable to the emission from HD~93129A, and therefore the estimated variability intrinsic to HD~93129A is unreliable.

The significant long-term variability detected is qualitatively consistent with the expected orbital modulation of the X-ray emission. The theoretical expectation for a thermal X-ray emission produced by the wind-wind interaction in adiabatic conditions is that the unabsorbed flux varies as $\propto 1/D$ \citep{Stevens1992}. However, testing this from observations requires to fit more complex spectral models, particularly in the soft X-ray band, which is the most affected by absorption. 
Moreover, converting the date of observation to system separation $D$ requires a precise knowledge of the orbital parameters. For this reason we did not aim to present a quantitative analysis of the flux variability in this work.

The spectrum above 12~keV is poorly constrained. Additional uncertainties arise given that \textit{NuSTAR} is unable to resolve the emission coming from a $30\arcsec$ region centred at the position of HD~93129A. Thus, it is impossible to determine the actual fraction of the observed flux that corresponds to HD~93129A. Nonetheless, \textit{Chandra}'s high angular resolution is capable of disentangling the different contributions up to 8~keV. We find that approximately half of the emission in the 3--8~keV range comes from the CWB. The spectral shape of the background sources is consistent with the one from HD~93129A. Moreover, the $\Gamma$ value we obtained from a power-law fit is consistent with the value of $\Gamma \lesssim 2$ found for $\eta$-Car by \citet{Hamaguchi2018}. This gives further support to the idea that the CWB HD~93129A is (partly) responsible for the observed hard X-ray emission. If we assume this extrapolation remains valid above 8~keV, we expect that roughly 50\% of the observed flux by \textit{NuSTAR} comes from HD~93129A as well. 
 
 Whether the measured 8--18~keV flux is an upper limit or a good estimate of the emission from HD~93129A depends on the assumptions made. 
 \begin{enumerate}
     \item  First, we can place a hard limit to the maximum flux for the power-law IC component by considering that none of the observed flux in the 8--18~keV range comes from HD~93129A. In this case we obtain $F_{8-18} \leq 2.6 \times 10^{-13}$\flux\, (considering the $1\sigma$ upper-limit). 
     \item A similar but less restrictive constraint is obtained by stating that the flux from the power-law IC component cannot be higher than the flux from the fitted power-law component. In this case we obtain $F_{8-18} \leq 2.2 \times 10^{-13}$\flux\, (again assuming the $1\sigma$ upper limit).
     \item The analysis in Sect.~\ref{sec:res-chandra} suggests that roughly $\sim 50\%$ of the observed hard X-ray flux comes from HD~93129A. In this case we can estimate the flux from the power law component as $F_{8-18} \approx 1.1 \times 10^{-13}$\flux.
 \end{enumerate}
The most conservative is the first.

Assuming the hard X-ray spectrum is thermal (with solar abundances), the maximum plasma temperature for an adiabatic shock in the WCR is $kT = 1.17 v_\mathrm{w,8}^2$~keV \citep{Stevens1992}, with $v_\mathrm{w,8}$ the wind velocity in units of $10^8$~cm~s$^{-1}$.
For a wind speed of $v_\mathrm{w,8} \approx 3$ \citep{Cohen2011}, maximum (i.e. close to the apex) shock temperatures $kT \approx 10$~keV are expected. This is within the poorly constrained values obtained in Sect.~\ref{sec:res-nustar}.
%


The CWB HD~93129A does not display a significant $\gamma$-ray activity as to be detected by \textit{AGILE}. 
We note that there is a low significance detection of emission near April 2018 at a 4-$\sigma$ level if emission from $\eta$-Car is not considered. Despite $\eta$-Car being a known $\gamma$-ray source located at just $\sim 0.2\degr$ from HD~93129A (i.e., within the same beam for \textit{AGILE}), during this epoch $\eta$-Car was expected to be in a low $\gamma$-ray emission state \citep{Balbo2017,White2019}. Thus, it is not likely that the detected flux of  $8\times10^{-7}$~cm$^{-2}$~s$^{-1}$ is due to $\eta$-Car. Nonetheless, it is less likely that it comes from HD~93129A, as its $\gamma$-ray flux was expected to be increasing at that epoch and no emission was found afterwards. Moreover, under the constraints imposed by not exceeding the observed hard X-ray flux, the $\gamma$-ray flux predicted by our non-thermal radiative model is $< 5\times10^{-12}$~cm$^{-2}$~s$^{-1}$, which is more than four orders of magnitude below the detected $\gamma$-ray flux (more details in Sect.~\ref{sec:modelling}). We also considered a possible hardening in the non-thermal electron distribution as in \citet{delPalacio2016}, which can enhance the $\gamma$-ray emission in a factor ten, but still it is not possible to account for the detected $\gamma$-ray flux even considering uncertainties in the system or model parameters. Finally, this emission is detected only at a 2-$\sigma$ level in the weekly-binned light-curves, and is thus not considered to be significant. We conclude that this emission is either a statistical fluctuation or due to a variable background. 

\subsection{Theoretical modelling} \label{sec:modelling}

We use an updated version of the non-thermal emission code presented in \citet{delPalacio2016} to calculate the predicted X-ray flux from HD~93129A. The modifications introduced in the model are described in Appendix~\ref{sec:NT_model}. We adopt similar parameters for the stellar winds; specifically, mass-loss rates $\dot{M}_1=10^{-5}$~M$_\odot$~yr$^{-1}$ and $\dot{M}_2 = 6\times10^{-6}$~M$_\odot$~yr$^{-1}$, and wind terminal 
velocities $v_{\infty,1}=3200$~km~s$^{-1}$ and $v_{\infty,2}=2800$~km~s$^{-1}$. This model has two free parameters: the ratio between the magnetic field pressure to thermal pressure in the WCR, $\eta_\mathrm{B}$, and the fraction of the available power at the shocks that is converted into relativistic electrons, $f_\mathrm{NT,e}$. The available power for particle acceleration is the wind kinetic power injected perpendicularly to the WCR, which is calculated consistently in the model. It is possible to tie these two parameters by modelling the synchrotron component revealed by radio data by \citet{Benaglia2015}. In this case the relation is $f_\mathrm{NT,e} B^2 = constant$ \citep{delPalacio2016}, which we assume holds along the orbit. However, it is not possible to break the degeneracy between these two parameters from radio data alone. 

\begin{figure}
\includegraphics[angle=270,width=1.00\linewidth]{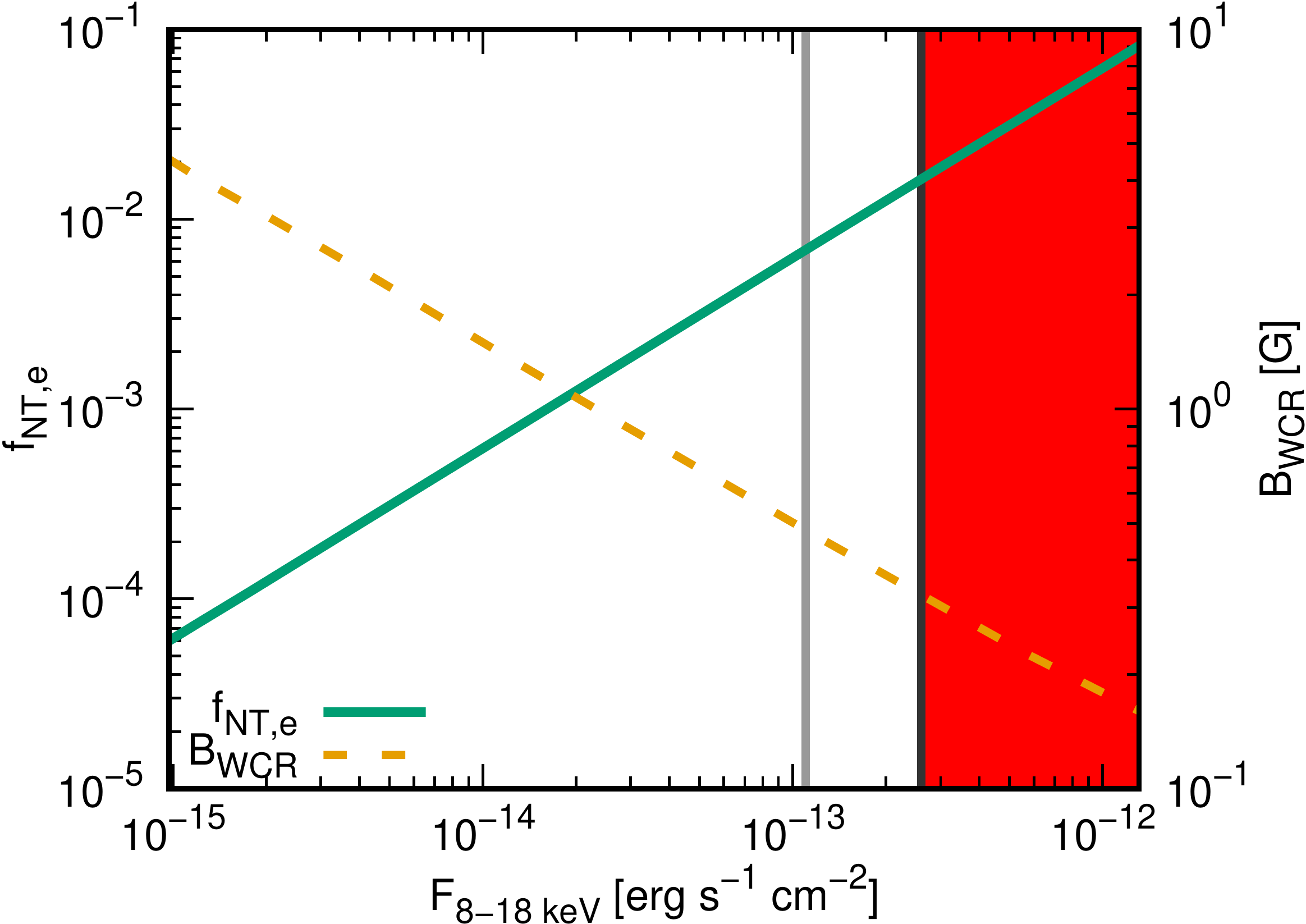}
\caption{Estimated values of $f_\mathrm{NT,e}$ and $B_\mathrm{WCR}$ in order to obtain a certain X-ray flux in the 8--18 keV. We mark with a dark grey line the strict (most conservative) upper limit for the non-thermal X-ray flux, which assumes that none of the flux observed with \textit{NuSTAR} comes from HD~93129A. 
The red region corresponds to values ruled out by the observations. We also show with a grey line the estimated flux value assuming that the relative contribution of HD~93129A to the X-ray flux with respect to the background remains constant above 8~keV (see Sect.~\ref{sec:obs_vs_predictions}).} 
\label{fig:FX_vs_B}
\end{figure}

Theoretical estimates of the non-thermal X-ray flux combined with the X-ray fluxes observed in the 8--18~keV allow us to constrain $f_\mathrm{NT,e}$. This is the only parameter that has a significant impact in determining the non-thermal X-ray flux, which is $F_X \propto f_\mathrm{NT,e}$. We adopt a periastron distance of $D = 19$~AU (see Sect.~\ref{intro}) and compute the model 8--18~keV flux for reasonable values of $f_\mathrm{NT,e}$. It is also possible to estimate the magnetic field in the apex of the WCR, $B_\mathrm{WCR}$, by assuming that the relation between $f_\mathrm{NT,e}$ and $\eta_\mathrm{B}$ holds along the orbit. A constant value of $\eta_\mathrm{B}$ is consistent with $B_\mathrm{WCR}$ scaling as $1/D$. This, in turn, is consistent with a constant (or null) $B$-field amplification factor in the WCR. Taking into account that $f_\mathrm{NT,e} B^2 = constant$ and $F_X \propto f_\mathrm{NT,e}$, we get the relation $B \propto F_X^{-1/2}$. In Fig.~\ref{fig:FX_vs_B} we show at the left vertical axis the required value of $f_\mathrm{NT,e}$ in order to reach a certain 8--18~keV flux, and the corresponding value of $B_\mathrm{WCR}$ at the right vertical axis.

\begin{figure}
\includegraphics[angle=270,width=1.0\linewidth]{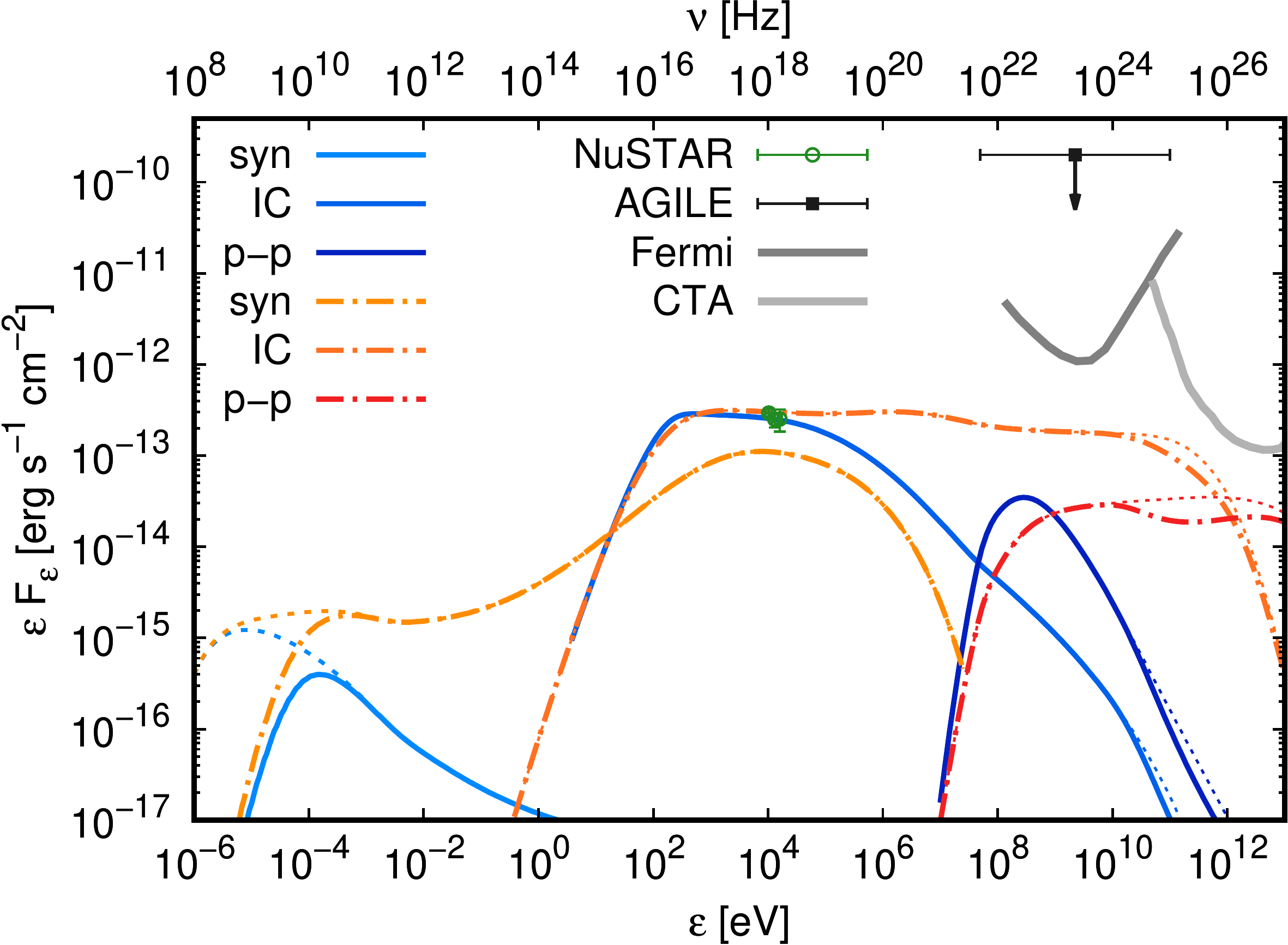}
\caption{Modelled SED adopting $f_\mathrm{NT,e} = 0.006$, $f_\mathrm{NT,p} = 10\;f_\mathrm{NT,e}$, and $B_\mathrm{WCR} \approx 0.5$~G. Solid lines are calculated for an injected particle energy distribution with constant spectral index of $p=3.2$, whereas dash-dotted lines consider a possible hardening of $p=2$ in the high-energy spectrum. Dotted lines show the intrinsic emission components before correcting for absorption effects (see Appendix~\ref{sec:NT_model} for details of the model). We also show the \textit{NuSTAR} data in the 8--18~keV energy range, the \textit{AGILE} upper-limit assuming a photon index $\Gamma = 2$, and the expected sensitivity for 4-yr \textit{Fermi}-LAT and for 100-h CTA \citep[extracted from][]{Funk2013}.} 
\label{fig:SED}
\end{figure}

We can interpret Fig.~\ref{fig:FX_vs_B} in two ways: (i) considering the upper limit to the power-law component in X-rays, from which we obtain an upper limit of $f_\mathrm{NT,e} < 0.02$ and a lower limit of $B_\mathrm{WCR} > 0.3$~G ($\eta_\mathrm{B} > 0.01$); and (ii) considering a detection at the estimated value of $F_{8-18}=1.1\times10^{-13}$\flux, from which we estimate $f_\mathrm{NT,e} \approx 0.006$ and $B_\mathrm{WCR} \approx 0.5$~G ($\eta_\mathrm{B} \approx 0.02$). For this latter case, we calculate the SED for two scenarios, one in which the injected particle energy distribution has a constant spectral index $p=3.2$, and one more favourable for $\gamma$-ray production in which the injected distribution hardens at high energies \citep[$> 100$~MeV for electrons; see][]{delPalacio2016}. In Fig.~\ref{fig:SED} we show the modelled SED together with the observational data for the periastron passage. Unfortunately, further constraints to the non-thermal particle population cannot be placed using the \textit{AGILE} upper limit as it is much higher than the $\gamma$-ray flux predicted in the most favourable scenario. In addition, as already discussed by \citet{delPalacio2016}, it is difficult to observe this system in the radio band close to periastron passage: despite the intrinsic synchrotron flux increases close to periastron as it scales with $B_\mathrm{WCR}$, the absorption of the low-frequency photons in the stellar winds also boosts during this epoch, resulting in a reduced flux below 10~GHz.
    
It is possible to relate the inferred magnetic field intensity in the WCR to that in the stellar surface. At large distances from the star ($r \gg R_\star$) the stellar magnetic field is toroidal and drops as $\propto~r^{-1}$. Following \citet[][and references therein]{delPalacio2016}, we can express $B_\star \approx 2.5 B_\mathrm{WCR} \, (r/R_\star)$. This expression considers an Alfv\'en radius $r_\mathrm{A} \sim R_\star$, a stellar rotation velocity $v_\mathrm{rot} \approx 0.1 v_\infty$, and also takes into account the adiabatic compression of magnetic field lines in the WCR.
We set $r$ equal to the periastron separation $D$ and allow, again, for two interpretations: (i) on the one hand, we obtain $B_\star > 130$~G for the lower limit of $B_\mathrm{WCR} > 0.3$~G; (ii) on the other hand, we get $B_\star \approx 200$~G for the estimated value of $B_\mathrm{WCR} \approx 0.5$~G. Such a modest $B$-field strength would not be sufficient to confine the winds (i.e. $r_A \leq R_\star$, as assumed). This is also consistent with insignificant emission of X-rays above $\sim 1$~keV by the individual stellar winds, which would require them to be magnetically confined \citep[e.g.][]{Ud-Doula2009}. 

We clarify that magnetic field amplification processes in the WCR \citep[e.g.][for studies in the context of supernova remnants]{Bell2004, Drury2012,delValle2016} were not considered nor computed in the calculations. If the measured magnetic field in the WCR actually results from some $B$-field amplification, this would translate into a stellar value lower than the one determined from the sole toroidal geometric dilution and adiabatic compression.

%
\section{Conclusions}\label{sec:conc}
%

We report the results from an observational campaign on the extreme colliding wind binary HD~93129A close to its periastron passage in 2018. The obtained optical data allowed us to determine the periastron epoch precisely (results will be presented in a forthcoming work), which we used to trigger our campaign in the high-energy domain. 
We observed HD~93129A with \textit{Chandra}, \textit{NuSTAR}, and \textit{AGILE}, covering a wide range of energies in the X-ray and $\gamma$-ray domain. We conclude that there is no significant $\gamma$-ray emission from this system and the upper limit of the flux in the 50~MeV--100~GeV range, $F < 3\times 10^{-7}$~cm$^{-2}$~s$^{-1}$, is unconstraining for non-thermal emission models. Our main observational result is the detection of emission in the hard X-ray band with \textit{NuSTAR} consistent with being (partially) produced by HD~93129A. It is not possible at this time to assess to what extent background contamination is accountable for this emission. Future observations of the source might reveal a persistent flux of the same value, which would indicate that this emission is produced by nearby sources, or a diminished flux, which would confirm HD~93129A as the responsible for this emission. 

We interpret the derived hard X-ray fluxes using the tightly constrained periastron distance from the optical monitoring and the non-thermal radiative model described in \citet{delPalacio2016}. This multi-zone model takes into account the relevant physical processes in the wind-collision region. The model has two parameters that can be constrained or estimated by our observations: the fraction of the wind kinetic power injected into the WCR that is converted into relativistic electron acceleration, $f_\mathrm{NT,e}$, and the magnetic field in the wind-collision region, $B_\mathrm{WCR}$. We present the conclusions for two different interpretations:
\begin{itemize}

    \item Under the very conservative assumption that none of the X-ray flux above 8~keV is produced by HD~93129A, we obtain $f_\mathrm{NT,e} < 0.02$. In addition, we estimate the magnetic field in the wind-collision region as $B_\mathrm{WCR} > 0.3$~G. Neglecting possible magnetic-field amplification in the wind-collision region, we derive a lower limit for the surface stellar magnetic field of $B_\star > 130$~G.

    \item We consider that $\sim 50\%$ of the 8--18~keV flux produced by a power-law component comes from HD~93129A. In this case we can estimate $f_\mathrm{NT,e} \approx 0.006$. In addition, we get $B_\mathrm{WCR} \approx 0.5$~G, from which we derive an upper limit for the surface stellar magnetic field $B_\star \leq 200$~G taking into account possible $B$-field amplification.

\end{itemize}
 
We conclude that multi-wavelength, dedicated observing campaigns during carefully selected epochs is a powerful tool for characterising the relativistic particle content and magnetic field intensity in colliding wind binaries. This, in turn, allows to constrain the value of the magnetic field on the surface of very massive stars. We also highlight the need for more sensitive and higher angular-resolution observations in the $\gamma$-ray band in order to better characterise the non-thermal particle population in colliding-wind binaries.

%

\section*{Acknowledgements}

We thank the anonymous referee for his/her comments that helped to improve the manuscript. S.d.P. thanks the Committee on Space Research (COSPAR) for support via its Capacity Building Program and its Capacity Building Fellowship Program. S.d.P., D.A. and F.G. acknowledge support from the Royal Society (RS) International Exchanges "The first step for High-Energy Astrophysics relations between Argentina and UK". F.G. acknowledges support from Athena project number 184.034.002, partly financed by the Dutch Research Council (NWO). D.A. acknowledges support from the RS. J.M.P., V.B.R. and G.E.R. acknowledge support by the Spanish Ministerio de Econom\'{\i}a, Industria y Competitividad (MINEICO/FEDER, UE) under grants AYA2016-76012-C3-1- P, MDM-2014-0369 of ICCUB (Unidad de Excelencia `Mar\'{\i}a de Maeztu') and the Catalan DEC grant 2017 SGR 643. G.E.R. was also supported by PIP 0338 (CONICET). J.M.A. acknowledges support from the Spanish Government Ministerio de Ciencia, Innovaci\'on y Universidades through grant PGC2018-095\,049-B-C22. R.H.B.S. acknowledges DIDULS Regular Project No. 18143. K.H. is supported by the Chandra grant GO8-19010A.



\bibliographystyle{mnras}
\bibliography{references}



\appendix

\section{Images}

\begin{figure*}
\centering
\includegraphics[width=0.88\linewidth]{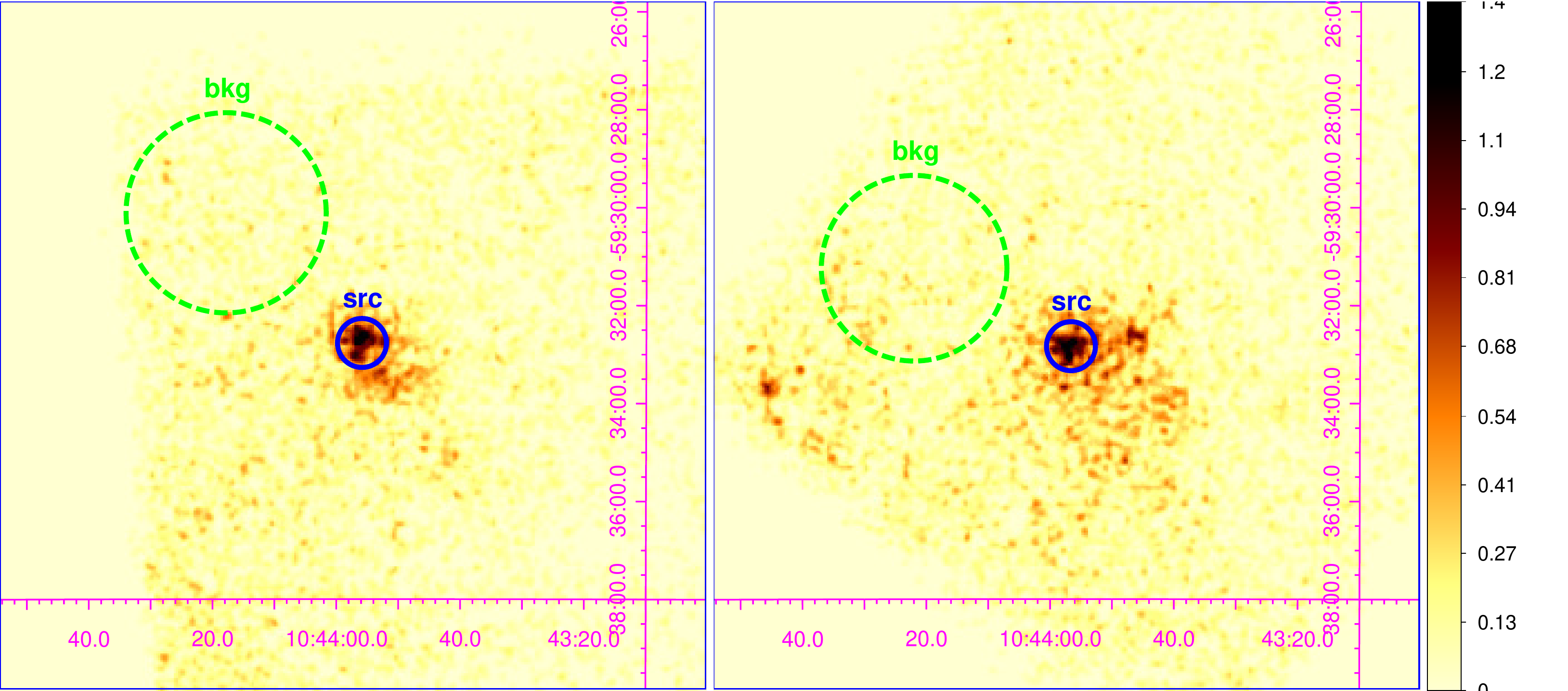}
\caption{\textit{NuSTAR} image with FPMB in the 3--11~keV energy range. The source and background extraction regions have a $30\arcsec$ and $2\arcmin$ radius, respectively. Left is the first observation, right is the second one.}
\label{fig:nustar_img}
\end{figure*}

We present here additional figures that complement our analysis. In Fig.~\ref{fig:nustar_img} we show a \textit{NuSTAR}-FPMB image of the field of view for the two observing epochs; FPMA image (not shown) is similar. The background extraction region was selected within the same chip that contained the source. Significant and variable background emission is observed close to the position of HD~93129A. 

\begin{figure}
\includegraphics[angle=270,width=1.0\linewidth]{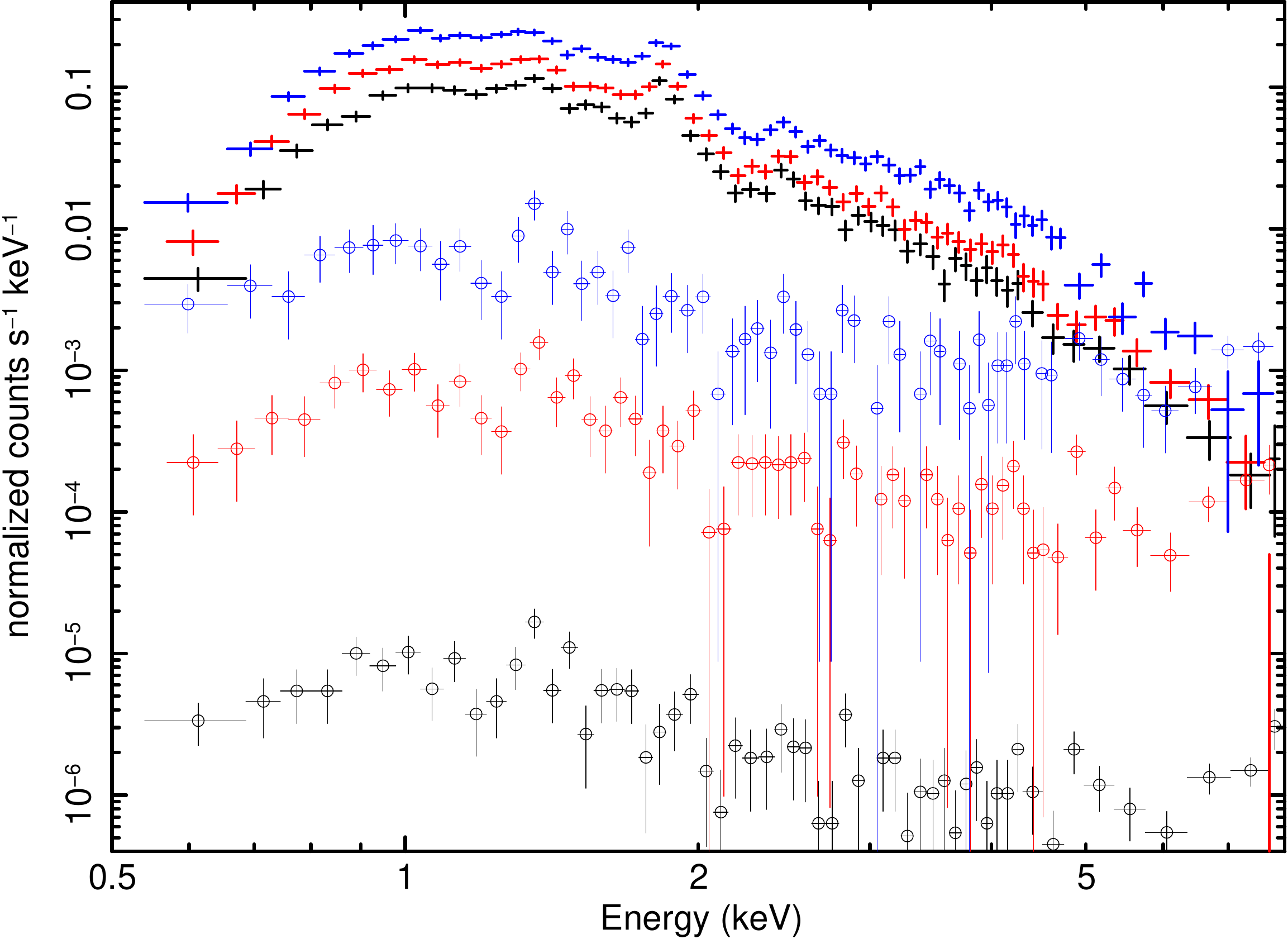}
\caption{Comparison of the \textit{Chandra} spectra extracted from a $1\arcsec$ (black), $10\arcsec$ (red), and $30\arcsec$ (blue) region centred at the position of HD~93129A. The crosses represent the source emission and the circles (with errorbars) the background. The spectra are rebinned (\texttt{setpl rebin 4 8} in \texttt{XSPEC}) for clarity.} 
\label{fig:chandra_1vs30}
\end{figure}

\begin{figure}
\includegraphics[angle=270,width=1.0\linewidth]{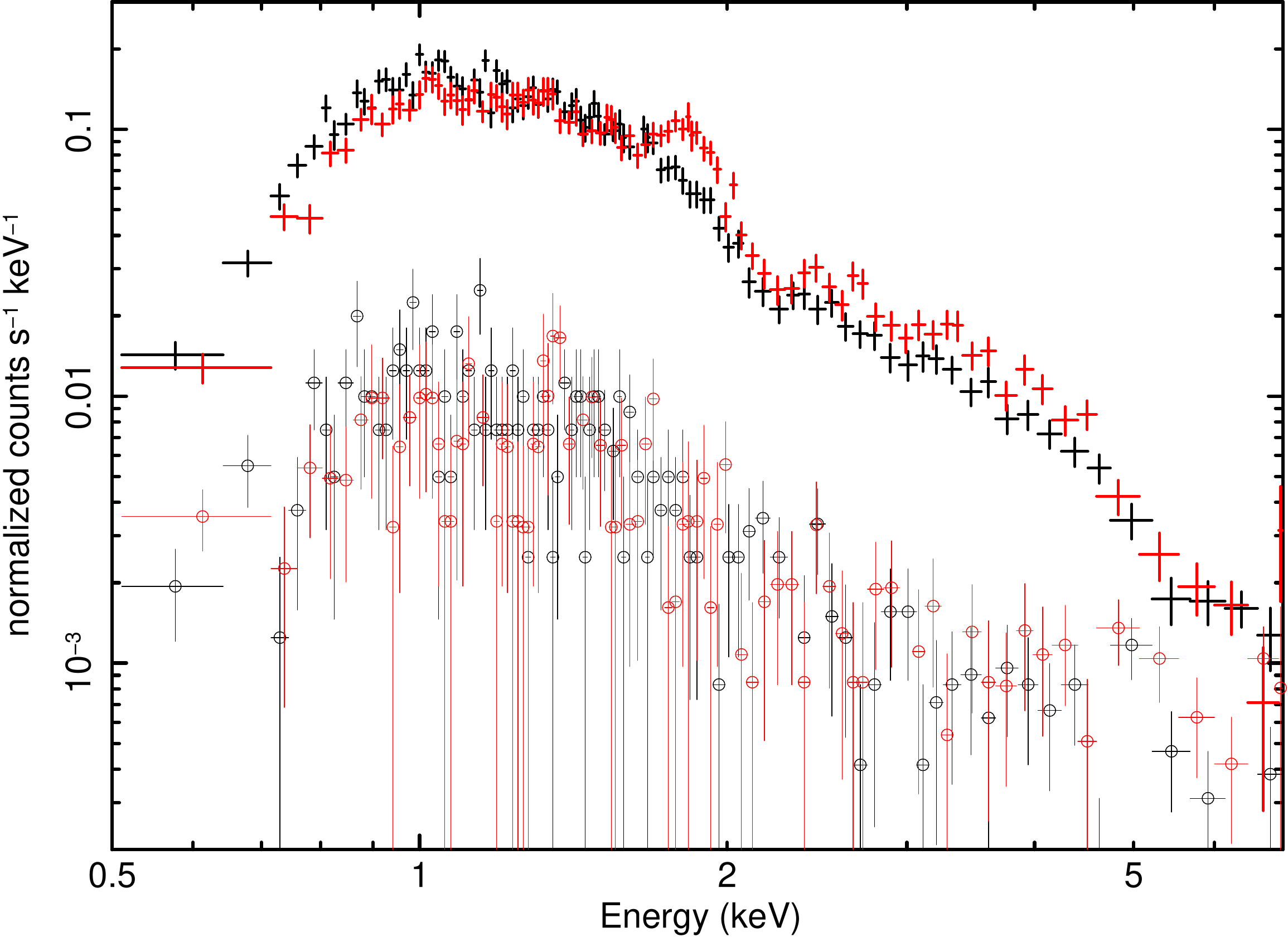}
\caption{Comparison of the \textit{Chandra} spectra of the sources within an annulus of outer radius $30\arcsec$ and inner radius $1\arcsec$ centred at the position of HD~93129A; black is 2004 and red 2018. The crosses represent the source emission and the circles the background.}
\label{fig:chandra_annulus}
\end{figure}

In Fig.~\ref{fig:chandra_1vs30} we show the comparison of the \textit{Chandra} source and background spectra for different source extraction regions of radii $1\arcsec$, $10\arcsec$, and $30\arcsec$. For the $1\arcsec$ source extraction region the background is negligible, whereas for radii $\geq 10\arcsec$ it becomes comparable to the source emission at energies above 7~keV. 
In addition, in Fig.~\ref{fig:chandra_annulus} we compare the combined \textit{Chandra} spectra of the nearby sources to HD~93129A between 2004 and 2018. The spectra were extracted from an annulus of outer radius $30\arcsec$ and an inner radius of $1\arcsec$ centred at the position of HD~93129A. The combined spectra from these sources are remarkably stable above 2 keV, where the continuum dominates. The apparent discrepancy below 2 keV is probably caused by a difference in the ACIS instrumental response: ACIS-S has a better soft band sensitivity than ACIS-I, but the soft band efficiency of all ACIS sensors has declined recently with contamination on the optical blocking filter. ACIS-S also shows a moderately strong feature around $E \lesssim 2$~keV. The flux difference between the two epochs is $< 5\%$ in the 0.5--3~keV range and $< 10\%$ in the 3--8~keV energy range.

\begin{figure}
\includegraphics[width=1.0\linewidth]{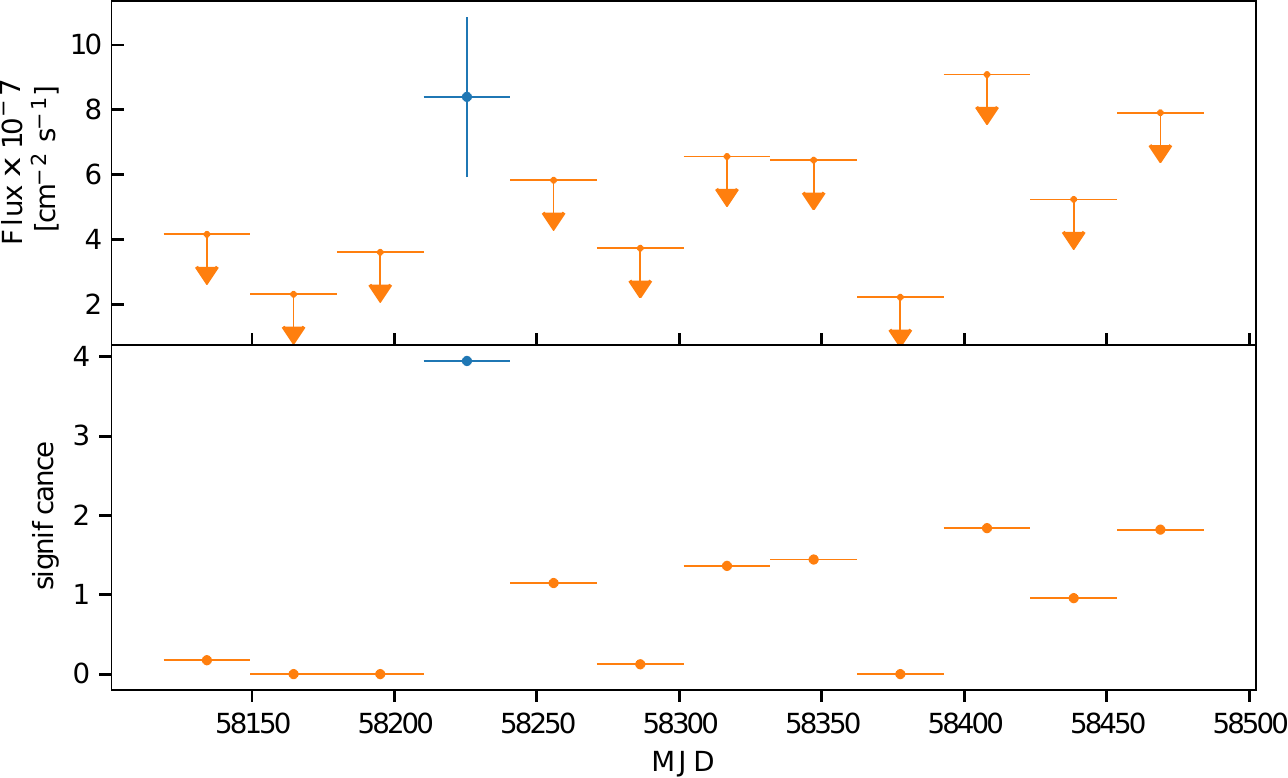}
\includegraphics[width=1.0\linewidth]{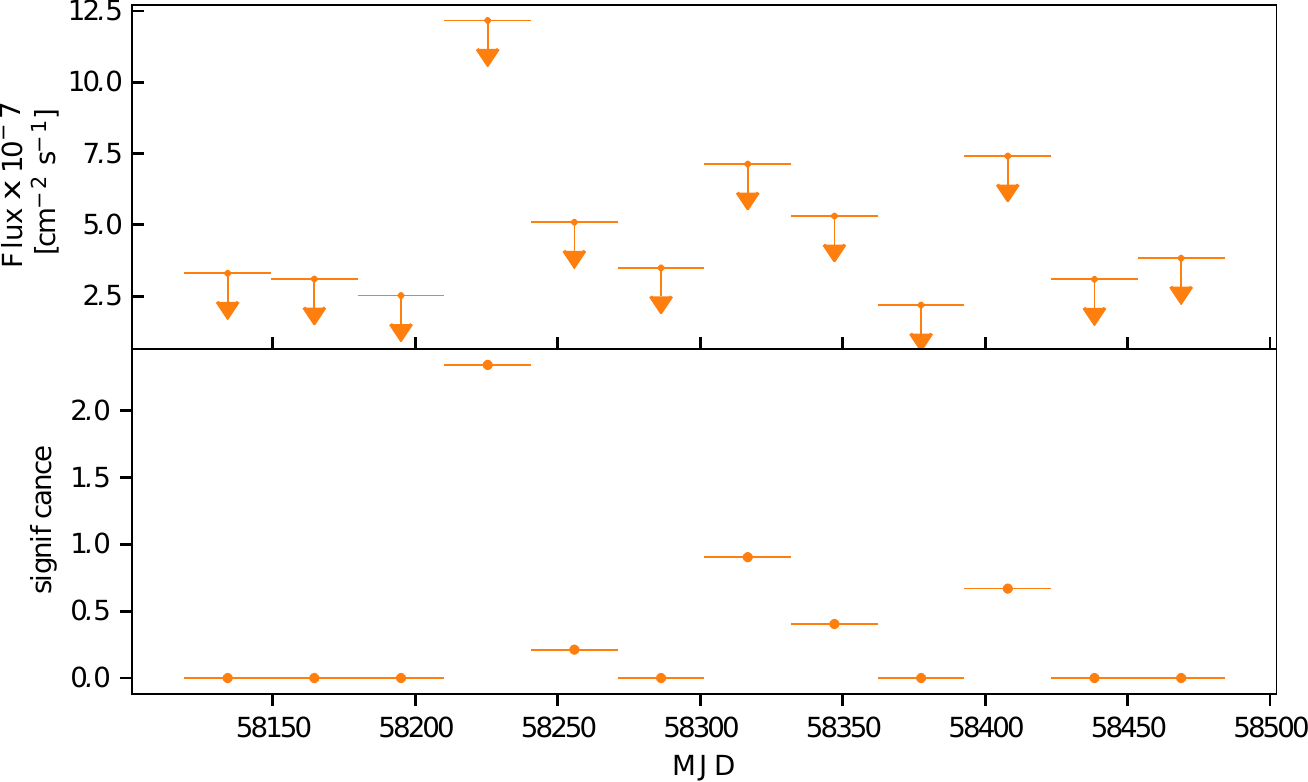}
\caption{\textit{AGILE} light-curve in the 50~MeV--100~GeV energy range during 2018. The top panel does not consider emission from $\eta$-Car, while the bottom panel considers it as an extra background source.}
\label{fig:agile}
\end{figure}

Finally, in Fig.~\ref{fig:agile} we present the monthly light curves obtained with \textit{AGILE} for two different background models. When the possible contribution from $\eta$-Car is not considered, a detection with a 4-$\sigma$ significance is obtained in April 2018. However, when $\eta$-Car is included as an additional background source the detection is not significant and only upper limits are obtained.

\section{Non-thermal emission model} \label{sec:NT_model}

Here we present a review of the non-thermal radiation model used. This model is an update to the one presented in \citet{delPalacio2016}, suitable for adiabatic and quasi-stationary shocks with a laminar flow. The WCR structure is treated as a 2-dimensional surface under a thin shock approximation. We assume that relativistic particles are accelerated once a fluid line from the stellar wind enters the WCR region. These particles flow together with the shocked fluid which convects the ambient magnetic field. As they stream, particles cool down due to different processes and produce broadband radiation. This emission is corrected for absorption by interaction with the local matter and radiation fields.

The relativistic particle distribution injected at a given position in the WCR is a power law with the spectral index given by the radio observations. The normalisation of this distribution is such that the injected power is a fraction $f_\mathrm{NT}$ of the total power available for particle acceleration \citep[which is only a fraction of the total power of the stellar winds;][]{delPalacio2016}. This power is distributed in electrons and protons as $f_\mathrm{NT} = f_\mathrm{NT,e} + f_\mathrm{NT,p}$. It is usual to define $f_\mathrm{NT,e} = K_\mathrm{e,p} f_\mathrm{NT}$, with  $K_\mathrm{e,p} \sim 0.01 - 0.1$ \citep[see][for a discussion of uncertainties in this value]{Merten2017}. For a canonical value of $f_\mathrm{NT} \sim 0.1$, we expect $f_\mathrm{NT,e} \sim 10^{-3}$, but with a large uncertainty.

The emission in the radio band is produced by synchrotron emission. This radiation can be significantly attenuated by free-free absorption in the ionised stellar winds. The non-thermal X-ray emission is produced by anisotropic inverse-Compton up-scattering of stellar photons. This process can dominate the $\gamma$-ray emission as well, competing with proton-proton inelastic collisions. The $\gamma$-ray photons can be absorbed in the stellar radiation field creating secondary electron-positron pairs. 

We introduced the following modifications to the model in \citet{delPalacio2016}:
\begin{itemize}
    \item The angle $\psi$ such that the observed distance  between the stars is $D_\mathrm{proj} = D \cos \psi$ is no longer a free parameter of the model\footnote{In \citet{delPalacio2016} this angle was referred as $i$, which can be confused with the orbital inclination.}. A value of $\psi \sim 32\degr$ is used in accordance with the most recent orbital ephemeris. 
    \item We considered an increased free-free opacity at radio-frequencies due to clumping in the stellar winds. This effect enhances the opacity by a factor $f^{-1/2}$, where $f\sim 0.1$ is the volume filling factor of the wind \citep{Muijres2011}. In this case there is no need to adopt a large value of $E_\mathrm{min,e}$ to reproduce the spectral break at low frequencies as done in \citet{delPalacio2016}. Instead, a typical value of $E_\mathrm{min,e} \approx 1$~MeV yields a good fit of the radio spectra.
    \item The IC spectra is calculated using the expressions by \citet{Khangulyan2014} suitable for black-body-like target photon fields. This reduces the computation time significantly.
    \item A small correction was introduced in the way particle evolution along stream lines is calculated. This considers variations of cooling times from cell to cell in the emitter \citep[Eq. 16 from][]{delPalacio2018}. 
\end{itemize}


\bsp	
\label{lastpage}
\end{document}